\documentclass[]{tPHM2e}
\usepackage{graphicx}

\def\bsym#1{\boldsymbol{#1}}
\def\eps{\varepsilon}
\def\gdir#1{\langle#1\rangle}

\def\refeq#1{(\ref{#1})}
\def\reffig#1{Fig.~\ref{#1}}

\articletype{}

\begin{document}

\title{Atomistic studies of transformation pathways and energetics in plutonium}

\author{R.~Gr\"oger$^{{\rm a}\ast}$\thanks{$^\ast$Corresponding author. Email: groger@lanl.gov}, 
  T.~Lookman$^{\rm a}$ and A.~Saxena$^{\rm a}$\\
  \vskip6pt $^{\rm a}$\em Theoretical Division, Los Alamos National Laboratory, Los Alamos, NM 87545, USA}

\maketitle

\begin{abstract}
  One of the most challenging problems in understanding the structural phase transformations in Pu
  is to determine the energetically favored, continuous atomic pathways from one crystal symmetry to
  another. This problem involves enumerating candidate pathways and studying their energetics to
  garner insight into instabilities and energy barriers. The purpose of this work is to investigate
  the energetics of two transformation pathways for the $\delta \rightarrow \alpha'$ transformation
  in Pu that were recently proposed \cite{lookman:08} on the basis of symmetry.  These pathways
  require the presence of either an intermediate hexagonal closed-packed (hcp) structure or a simple
  hexagonal (sh) structure. A subgroup of the parent fcc and the intermediate hexagonal structure,
  which has trigonal symmetry, facilitates the transformation to the intermediate hcp or sh
  structure. Phonons then break the translational symmetry from the intermediate hcp or sh structure
  to the final monoclinic symmetry of the $\alpha'$ structure. We perform simulations using the
  modified embedded atom method (MEAM) for Pu to investigate these candidate pathways. Our main
  conclusion is that the path via hcp is energetically favored and the volume change for both
  pathways essentially occurs in the second step of the transformation, i.e. from the intermediate
  sh or hcp to the monoclinic structure. Our work also highlights the deficiency of the current
  state-of-the-art MEAM potential in capturing the anisotropy associated with the lower symmetry
  monoclinic structure.
\end{abstract}


%
%

\section{Introduction}

One of the outstanding problems in Pu science is understanding the mechanism of the structural
transformation from the high-temperature fcc $\delta$ phase to the low-temperature monoclinic
$\alpha$ phase \cite{hecker:00, hecker:04, schwartz:07, lookman:08}. This transformation is
accompanied by changes in the shape of the unit cell and numbers of atoms within the unit
cells. Thus, a combination of strains and shuffles (displacement modes) are involved in describing
these transformations and the challenging problem is to obtain the atomic pathways from one crystal
symmetry to another. This is a highly nontrivial problem as the nature of the transformation
mechanism itself depends crucially on orientation relationships between the structures and,
therefore, its elucidation requires exploring a myriad of symmetry relationships between them. For
example, a reconstructive transformation typically involves the presence of intermediate structures
not related to the parent and the product phases by group-subgroup relations that allow the atoms to
move continuously, subjected to the constraints imposed by the orientation relationships.  In
contrast, a displacive transformation is a one-step group-subgroup process occurring without the
need to invoke intermediate structures and involves relatively straightforward orientation
relations.  Recent advances in exploring and searching through large databases of crystal symmetries
\cite{aroyo:06, aroyo:06a} have made this problem more tractable and provide the means of
enumerating candidate pathways the energetics of which can be subsequently studied by electronic
structure or molecular dynamics calculations.

\begin{figure}[htb]
  \centering
  \footnotesize
  \includegraphics[scale=0.7]{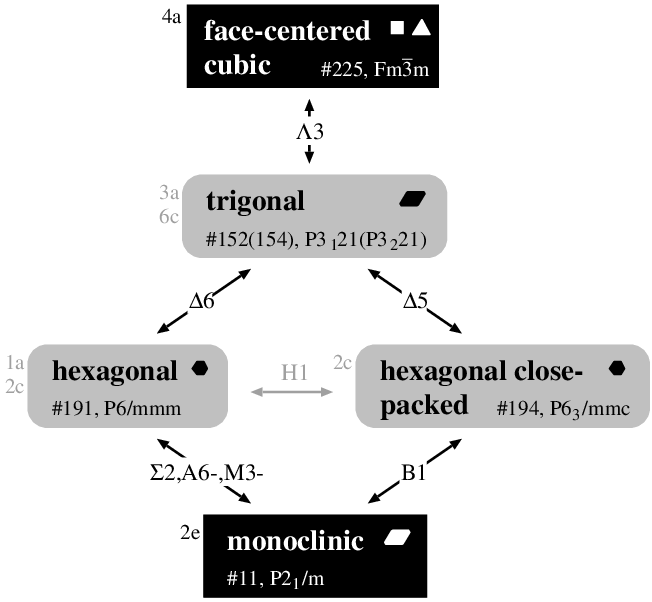} \\
  \caption{The phonon mechanism for $\delta \rightarrow \alpha(\alpha')$ phase transformation as
    proposed in Ref.~\cite{lookman:08}. The arrows indicate displacive (martensitic) phase
    transformations between structures that are group-subgroup related. The labels attached to
    individual arrows indicate specific phonons and elastic constants associated with the strains of
    the parent structure that drive the corresponding transformation. The Wyckoff symbols of each
    structure are attached to the left of each box.}
  \label{fig_delta-alpha}
\end{figure}

Lookman et al. \cite{lookman:08} have recently suggested that the transformation from the fcc
$\delta$ phase stabilized by certain trivalent impurities, such as Ga, Al, Ce or Am, to the
monoclinic $\alpha'$ phase can be accomplished using specific strains, displacement modes, and via
intermediate structures. This model is consistent with experimental data, notably the results of
phonon dispersion experiments obtained by inelastic X-ray scattering on a thin polycrystalline
sample of Pu-Ga alloy \cite{wong:03}. The presence of these impurities merely changes the lattice
parameters and local ordering but does not affect the global crystal symmetry, hence $\delta$ and
(Ga, Al, Ce, Am)-stabilized $\delta$, as well as $\alpha$ and $\alpha'$, are assumed to have the
same respective crystal symmetries.  On the basis of symmetry, the experimental orientation
relationships, i.e. $[111]_\delta$ $\parallel$ $[010]_\alpha$ and $[1{\bar 1}0]_\delta$ $\parallel$
$[11{\bar 2}]_\alpha$, impose severe constraints on the $\delta \rightarrow \alpha(\alpha')$
transformation. As a consequence, the required intermediate structure turns out to possess hexagonal
symmetry and the transformation proceeds in two steps. The first step is from the fcc structure to
the hexagonal symmetry facilitated by their common subgroup, the trigonal symmetry. The second is
from the hexagonal to a monoclinic structure, which is a group-subgroup transformation driven by
collective displacement modes, or phonons, that break translational symmetry by combining either 8
hexagonal close-packed (hcp) or 16 simple hexagonal (sh) primitive cells. \reffig{fig_delta-alpha}
shows both transformation pathways via the common trigonal subgroup. The conclusion that the
intermediate crystal structure has to have hexagonal symmetry can be appreciated by recognizing that
the only way an fcc threefold $[111]$ axis can become a monoclinic twofold $[010]$ axis is by an
increase of symmetry via a structure with a sixfold axis (product of the individual symmetries of
the fcc $[111]$ and the monoclinic $[010]$ axes), that is characteristic for the hexagonal symmetry.

Although symmetry is immensely useful in constraining the enormous phase space of possible pathways,
it needs to be complemented by energy calculations to indicate the presence of instabilities and
barriers in order to determine the likely atomic pathways. First-principles electronic structure and
atomistic calculations of strongly correlated materials, in particular Pu and its alloys, have long
been at the forefront of computational approaches to complement scarcely available experimental data
and to provide insight into the fundamental physics. The density functional theory (DFT) in the
usual local density approximation (LDA) or the generalized gradient approximation (GGA) predicts a
30\% larger volume of the $\delta$ phase and magnetic long-range order. Electronic structure
calculations of Pu based on the full-potential linear augmented plane-wave (FP-LAPW) method
utilizing an antiferromagnetic configuration \cite{robert:03} have shown to reproduce the bulk
modulus of the $\alpha$, $\gamma$ and $\delta$ phases better than a non-spin-polarized
calculation. Currently the only first principles approach that has been demonstrated to reproduce
all solid-state phases of Pu utilizes the full-potential linear muffin-tin orbital (FP-LMTO) method
with spin-orbit coupling and orbital polarization effects \cite{soderlind:04}. In this work,
antiferromagnetic optimized configurations were used for the $\alpha$, $\beta$ and $\gamma$ phases,
while an approximation using disordered magnetic moments was adopted for $\delta$, $\delta'$ and
$\epsilon$ phases. At present, experiments suggest that there are no static or dynamic magnetic
moments in the $\alpha$ or $\delta$ phases of Pu \cite{lashley:05}, which may be due to complex
screening effects that obscure the experimental observations \cite{soderlind:04}.  The most recent
method of incorporating strong correlations arising from the almost half-filled $5f$ orbitals
\cite{albers:07} is based on augmenting DFT with the Hubbard Coulomb correlations between electrons,
giving rise to the ``DFT+U'' method. However, even with this additional complexity, the many-body
wavefunctions for a crystalline solid do not reduce to the corresponding wavefunctions for isolated
atoms as the lattice spacing is increased \cite{savrasov:01}. This drawback is removed in the more
recent dynamical mean field theory (DMFT) in which the Anderson impurity model \cite{cooper:00} is
utilized to study the tendency of the $f$ electrons to delocalize \cite{savrasov:01}.

Despite the success of DFT calculations at capturing correctly the internal energies of all crystal
structures \cite{soderlind:04} and the ability of DMFT to reproduce the phonon properties of the
$\delta$ phase of Pu \cite{dai:03}, they are so far impractical as a means to determine energies
for transformation pathways between the $\delta$ and $\alpha(\alpha')$ phases. These simulations
could be performed efficiently utilizing a well-parametrized empirical or semi-empirical interatomic
potential that would capture at least qualitatively the strong directional bonding arising from
itinerant behavior of the $5f$ electrons in the low-symmetry monoclinic $\alpha$ and $\beta$ phases
of Pu, and their localization in the structures of high symmetry such as the orthorhombic $\gamma$,
face-centered cubic (fcc) $\delta$ or body-centered cubic (bcc) $\eps$ structures. The current
state-of-the-art among these approaches is the modified embedded atom method (MEAM), developed
specifically for Pu by Baskes \cite{baskes:00}, in which each atom is embedded in the background
electron density formed by the $s$, $p$, $d$, and $f$ electron densities centered on all other
atoms. 

The purpose of this work is to formulate a simplified approach that allows studies of the energetics
of the two transformation pathways proposed in \cite{lookman:08} for the $\delta \rightarrow
\alpha'$, i.e. fcc to monoclinic, transformation via the intermediate hcp or simple hexagonal
structures.  We utilize atomistic simulations within the MEAM to probe the two candidate paths and
calculate the energetics associated with these paths. Our principal finding is that the $\delta
\rightarrow \alpha(\alpha')$ pathway via the intermediate hcp structure has lower energy barrier
than that obtained for the intermediate sh structure. Moreover, for both cases the volume change
principally occurs in the second step, that is, during the group-subgroup displacive transformation
from the hexagonal to the monoclinic symmetry. Our findings also point to serious deficiencies in
the current state-of-the-art MEAM potential for Pu which are presumably due to its inability to
reproduce the experimental positions of atoms in the low-symmetry $\alpha$ and $\beta$ phases of
Pu. We conclude by outlining possible improvements of these calculations that may allow for future
systematic search for transformation pathways and energetics without a priori specifying the space
group of the intermediate crystal structure.

\section{Interatomic potential and equilibrium crystal structures}

In the following calculations, we utilize the modified embedded atom method (MEAM) interatomic
potential for Pu \cite{baskes:00}, within which the total energy is written as
\begin{equation}
  E = E_{emb} + E_{pair} \ .
\end{equation}
Here, $E_{emb}$ is an energy to embed an isolated atom into a background electron density formed by
electrons centered on all other atoms, and $E_{pair}$ is a pair potential. The embedding energy
takes a form that is often used for the exchange-correlation functional in the LDA approximation of
the DFT. The adjustable parameters in the functional form of the embedding energy are calculated so
as the potential reproduces some fundamental properties of the fcc $\delta$ phase, e.g. the
experimental elastic constants. The pair potential is then constructed so as the dependence of the
total energy on volume for the reference fcc structure matches the universal equation of state
\cite{rose:84} that is parametrized by the experimental bulk modulus of the $\delta$ phase and by
the pressure derivative of the bulk modulus of the $\alpha$ phase\footnote{Similar measurements for
  the $\delta$ phase were not available at the time. These experiments were performed recently and
  are published in Ref.~\cite{faure:06}.}. If the theoretical formalism of this potential captures
the essential physical aspects of bonding at least qualitatively correctly, one might expect that
the properties associated with the $\delta$ phase and other high-symmetry structures that do not
deviate significantly from this reference lattice will be reproduced as well. On the other hand, the
monoclinic $\alpha$ phase has 20\% lower volume than the reference fcc structure and the potential
may not be capable to reproduce the experimental measurements with sufficient accuracy.

In order for our simulations to be consistent with the interatomic potential used, it is necessary
to obtain the positions of atoms and the lattice parameters of the monoclinic cell that corresponds
to the minimum of energy. This is accomplished by utilizing the isothermal-isobaric molecular
dynamics simulation at 0~K where the temperature is maintained constant by simple rescaling of
velocities to obey the equipartition of kinetic energy and the pressure is controlled by the
modularly invariant Cleveland-Wentzcovitch barostat \cite{cleveland:88, wentzcovitch:91}. The
latter allows the simulated cell to change shape as the atomic positions are evolved during the
minimization of energy. The ensuing system of coupled equations of motion for the atomic positions
and for the shape of the simulated cell (i.e. lattice parameters $a$, $b$, $c$ and the lattice
angles $\alpha$, $\beta$, $\gamma$) is integrated numerically using the fourth order Gear
predictor-corrector method \cite{allen:87} with constant time step. To speed up the relaxation
process we constrain the atoms to obey the twofold symmetry of the $\bsym{b}$ axis of the monoclinic
lattice. In particular, the equations of motion for the atoms are solved only for the eight
nonequivalent atoms in one of the two monoclinic planes while the positions of the other eight atoms
in the second plane are calculated from the symmetry mapping $(a,\frac{1}{4},c) \leftrightarrow
(\bar{a},\frac{3}{4},\bar{c})$, where $a$ and $c$ are fractional coordinates of atoms measured
relative to the $\bsym{a}$ and $\bsym{c}$ axes. The energy of the relaxed monoclinic lattice is
about $-4.1$~eV/atom which is in a good agreement with the values extrapolated from experimental
data \cite{baskes:00}. Similarly, the energy difference between the $\delta$ and $\alpha$
structures used in our simulations is about $0.3~$eV/atom, which agrees with the total internal
energy differences estimated in \cite{wallace:98} by accounting for the thermal expansion of the
lattice within each phase. The calculated volume of the $\alpha$ structure is 20.8~\AA$^3$/atom
which slightly overestimates the measured experimental volumes extrapolated to 0~K.

\section{Energetics of transformation pathways}

In the most general situation the prominence of individual phonon modes varies along the
transformation pathway, and this necessarily gives rise to the motion of atoms along curvilinear
paths. Unless one performs a more sophisticated simulations, such as the nudged elastic band
(NEB) method \cite{jonsson:98}, which cannot be done at present due to apparent difficulties of the
MEAM potential to reproduce the positions of atoms in the $\alpha$ structure (more on this later),
the information about the prominence of individual phonon modes is not available. Hence, one is
forced to assume that the superposition of a number of phonons responsible for a given
transformation results in the motion of atoms along the shortest paths connecting their initial and
final positions, which is a simplification that we also adopt in our simulations here. Although
these calculations were carried out for pure Pu, our conclusions are qualitatively valid also for
dilute alloys of Pu with Ga, Al, Ce or Am that stabilize the fcc $\delta$ phase down to essentially
0~K. The reason is that low concentrations of these impurities do not change significantly the
lattice parameters. In particular, additions of Ga and Al slightly shrink the fcc lattice while Ce
and Am cause its expansion \cite{hecker:00}. Nevertheless, for low concentrations of these alloying
elements the changes in the lattice constant do not exceed 1\% of the values for pure Pu.

\subsection{fcc $\rightarrow$ sh $\rightarrow$ monoclinic transformation}

\begin{figure}[htb]
  \centering
  \footnotesize
  \begin{minipage}{.4\textwidth}
    \centering
    fcc $\rightarrow$ sh \\
    \includegraphics{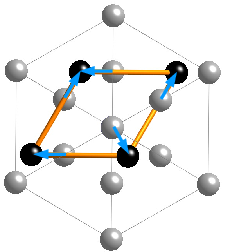} \\
    (a)
  \end{minipage}
  \begin{minipage}{.4\textwidth}
    \centering
    sh $\rightarrow$ monoclinic \\
    \includegraphics{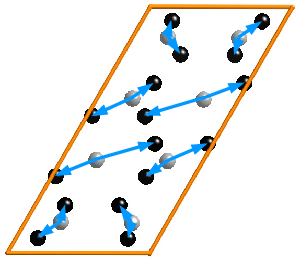} \\
    (b)
  \end{minipage} 
  \vskip1em
  \begin{minipage}{.4\textwidth}
    \centering
    fcc $\rightarrow$ hcp \\
    \includegraphics{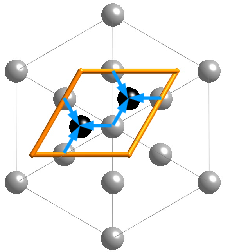} \\
    (c)
  \end{minipage}
  \begin{minipage}{.4\textwidth}
    \centering
    hcp $\rightarrow$ monoclinic \\
    \includegraphics{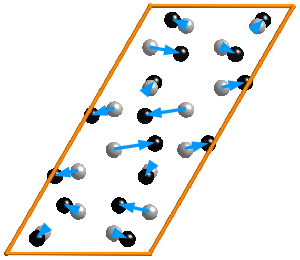} \\
    (d)
  \end{minipage}
  \caption{Motion of atoms during the fcc $\rightarrow$ sh (a), sh $\rightarrow$ monoclinic (b), fcc
    $\rightarrow$ hcp (c), and hcp $\rightarrow$ monoclinic (d) transformations. The atoms in the
    parent lattice (outlined) are plotted in gray and their corresponding positions after the
    transformation is complete are in black; the shape of the final cell is not shown for
    clarity. The arrows show the motion of the atoms between the two structures, i.e. the vectors
    $\hat{\bsym{t}}_i$ in \refeq{eq_rdispl}.}
  \label{fig_fcc-mcl_atoms}
\end{figure}

\begin{figure}[htb]
  \centering
  \footnotesize
  \includegraphics[scale=0.4]{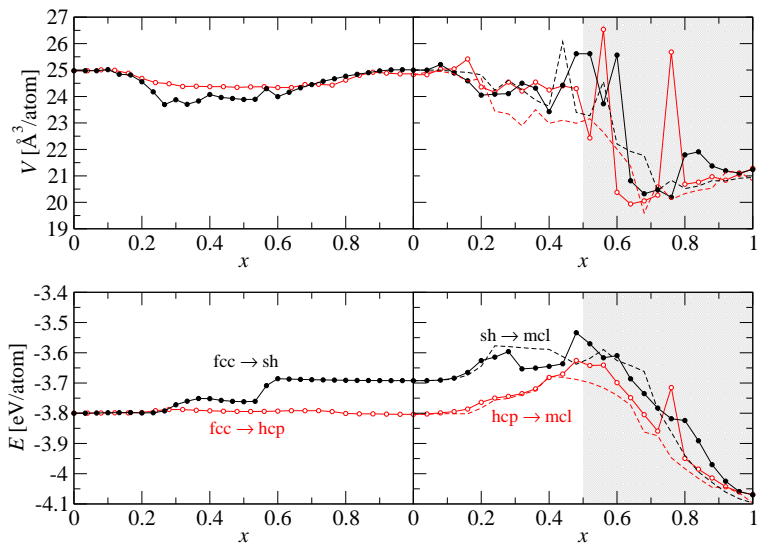}
  \caption{Variation of volume ($V$) and energy ($E$) during the transformation of Pu from its fcc
    $\delta$ phase to the monoclinic $\alpha$ phase. In the left panel of the figure the atoms are
    moving from their initial positions in the fcc lattice to their final positions in the sh (hcp)
    lattice, while in the right panel of the figure the atoms are moving from their positions in the
    sh (hcp) lattice to their final positions in the monoclinic lattice. The transformation
    coordinate $x$ designates the progress of the transformation ($x=0$ corresponds to the initial
    structure and $x=1$ to the final structure). In the right panel of the figure, the solid lines
    correspond to the 16-atom conventional monoclinic cell, that is stable within MEAM, with angle
    $106.3^\circ$ and the dashed lines to the 32-atom monoclinic supercell with angle
    $120.3^\circ$.}
  \label{fig_esurf}
\end{figure}

The simulated block is constructed by placing the atoms in their equilibrium lattice positions in
the fcc structure, as shown in \reffig{fig_fcc-mcl_atoms}(a). Due to the orientation relationship
[111]$_\delta$ $\parallel$ [0001]$_{\rm sh}$, the minimum number of atoms in the initial fcc
structure is three. During the course of the transformation, these atoms are considered to move
along the shortest paths from their initial positions, depicted as gray in
\reffig{fig_fcc-mcl_atoms}(a), to the three corners of the final hexagonal lattice in the directions
shown in the same figure by arrows. The extent of this transformation is conveniently measured by
the transformation coordinate $x$ that represents the displacement of each atom from its initial
(fcc) to the final (sh) lattice. Here, $x=0$ corresponds to the fcc lattice, $0 < x < 1$ to the
intermediate trigonal lattice, and $x=1$ to the final sh lattice. Each value of $x$ thus corresponds
to different fractional coordinates of atoms that are calculated as
\begin{equation}
  \hat{\bsym{r}}_i(x) = \hat{\bsym{r}}_i^0 + x\hat{\bsym{t}_i} \ ,
  \label{eq_rdispl}
\end{equation}
where the hat ($\,\hat{ }\,$) designates fractional coordinates measured relative to the shape of the simulated
cell, and $\hat{\bsym{t}}_i$ is the vector that displaces the atom $i$ from its initial position in
the fcc structure to its final position in the sh lattice. For known positions of atoms, given by
$\hat{\bsym{r}}_i(x)$, we calculate the shape of the unit cell that minimizes the total energy of
the system. During this optimization the fractional coordinates of atoms do not change and the
absolute motion of atoms is only due to the changing shape of the simulated cell. The relaxed
configuration is characterized by the lattice parameters $a$, $b$, $c$, the lattice angles $\alpha$,
$\beta$, $\gamma$, the corresponding volume of the cell $V$ and the energy $E$. The variation of the
volume and energy as a function of the transformation coordinate $x$ is shown in the left panel of
\reffig{fig_esurf}. The activation barrier for this transformation is about 0.1~eV/atom. An
important aspect to note is that no volume change is observed after the transformation is
completed. If the sh phase were the intermediate phase, it would thus mean that the 20\% volume
decrease during the $\delta$ $\rightarrow$ $\alpha(\alpha')$ transformation observed in experiments
has to occur during the sh $\rightarrow$ monoclinic transformation.

For the calculation of the sh $\rightarrow$ monoclinic transformation we utilized the relaxed sh
lattice from above to construct a $2 \times 2 \times 4$ sh supercell. The positions of atoms in this
initial supercell are shown in \reffig{fig_fcc-mcl_atoms}(b) as gray while their final positions in
the monoclinic lattice are in black. Although the assignment of atomic correspondences between the
two crystal structures (i.e. sh and monoclinic) is not unique, one can expect that the collective
motion of the atoms proceeds in such a way that the activation energy for this transformation is
minimized. The corresponding atomic correspondences are shown in \reffig{fig_fcc-mcl_atoms}(b) by
arrows that designate the direction of motion of each atom and thus the vector $\hat{\bsym{t}}_i$.
From the Shoji-Nishiyama relationship, the [0001] axis of this sh cell will coincide at the end of
the transformation with the [010] axis of the final monoclinic structure. This helps us to identify
the correspondences between the positions of atoms in the initial and final crystal structures that
are determined from the assumption that each atom in the initial sh supercell moves to its
corresponding position in the monoclinic cell in the shortest distance when measured in fractional
coordinates. The transformation coordinate $x$ now represents the displacement of atoms from their
initial position in the sh supercell, i.e. $x=0$ corresponds to the initial sh lattice and $x=1$ to
the final monoclinic lattice. The instantaneous position of each atom is again calculated using
\refeq{eq_rdispl} with the displacement vectors $\hat{\bsym{t}}_i$ indicated in
\reffig{fig_fcc-mcl_atoms}(b).

The calculated dependence of the relaxed volume of the cell and the corresponding energy is shown in
the right panel of \reffig{fig_esurf}. As expected above, the volume necessarily undergoes about
20\% decrease. The energy increases by an additional 0.2~eV/atom and then decreases to the final
value of about -4.1~eV/atom which is the energy of the monoclinic lattice that is stable within
MEAM. Accounting for the energy barriers from the two transformation steps, the total energy barrier
for the fcc $\rightarrow$ sh $\rightarrow$ monoclinic transformation becomes about 0.3~eV/atom.

\subsection{fcc $\rightarrow$ hcp $\rightarrow$ monoclinic transformation}

For this simulation the initial block of atoms contains six atoms in the six consecutive $(111)$
planes of the fcc lattice. The positions of atoms in the initial fcc lattice are shown in
\reffig{fig_fcc-mcl_atoms}(c) in gray while their corresponding positions in the final hcp lattice
are in black; the arrows indicate the direction of motion of each atom. After the atoms are
displaced in straight paths towards their corresponding positions in the hcp structure, where the
instantaneous positions of atoms are calculated from \refeq{eq_rdispl}, one obtains three hcp cells
stacked on the top of each other in the [0001] direction. The calculated dependence of the volume
and energy on the transformation coordinate $x$ is shown in the left panel of \reffig{fig_esurf}. As
in the case of fcc $\rightarrow$ sh step discussed above, there is no volume change after completing
this transformation and thus it is inevitable that the 20\% volume decrease occurs in the second
step, i.e. during the hcp $\rightarrow$ monoclinic transformation. Interestingly, there is a very
low energy barrier for the fcc $\rightarrow$ hcp transformation which implies comparable stability
of the fcc and hcp phases of Pu.

In order to investigate the hcp $\rightarrow$ monoclinic transformation, we again construct the
initial hcp supercell that now consists of a $2 \times 1 \times 4 $ array of the hcp unit cells
with two atoms per cell. The initial positions of atoms are shown in \reffig{fig_fcc-mcl_atoms}(d) in
gray and their final positions in the monoclinic lattice are in black. The arrows indicate the
displacement vectors $\hat{\bsym{t}}_i$ that are substituted in \refeq{eq_rdispl} to calculate
$\hat{\bsym{r}}_i(x)$, the instantaneous fractional coordinates of atoms corresponding to a given
value of the transformation coordinate $x$. The evolution of the volume and energy of the simulated
cell is shown in the right panel of \reffig{fig_esurf}. One again observes the 20\% volume decrease
and an additional energy barrier of about 0.2~eV/atom. However, since there is virtually no energy
barrier observed in the first step of this transformation, the total energy barrier for the fcc
$\rightarrow$ hcp $\rightarrow$ monoclinic transformation is only about 0.2~eV/atom and thus lower
than for the case when the intermediate phase has sh structure.

\section{Discussion}
\label{sec_discussion}

The results of the simulations presented in \reffig{fig_esurf} imply that the preferred intermediate
phase for the fcc ($\delta$) $\rightarrow$ monoclinic ($\alpha,\alpha'$) transformation in Pu is the
hcp phase and the corresponding energy barrier is about 0.2~eV/atom. It is well-known that Pu has a
unique position among the actinides that stems from the behavior of its $f$ electrons
\cite{lander:03, shim:07}. In particular, the $f$ electrons in early actinides (Ac to Np) are
itinerant and participate in bonding, whereas in the late actinides (Pu to No) they are localized at
individual atoms. Since the late actinides (in particular Am to Cf) all crystallize in hcp (or
double-hcp) structures and owing to the fact that Pu technically belongs to the group of late
actinides, it is not unreasonable to imagine that the hcp structure should be an intermediate
metastable phase mediating the transformation from the fcc to the monoclinic phase. An hcp phase has
been recently observed during the pressure-induced $\delta$ $\rightarrow$ $\alpha'$ transformation
in Pu-2at.\% Ga alloys \cite{faure:06}. Utilizing Rietveld refinement, the crystal structure of this
intermediate phase was found to belong to the orthorhombic Fddd group with the atomic positions
corresponding to those in slightly distorted hcp planes. This intermediate phase is presumably
short-lived (the duration of its stability is of the order of picoseconds) and, therefore, further
transient X-ray diffraction experiments are needed to confirm its existence.

\begin{figure}[htb]
  \centering
  \footnotesize
  \includegraphics[scale=0.3]{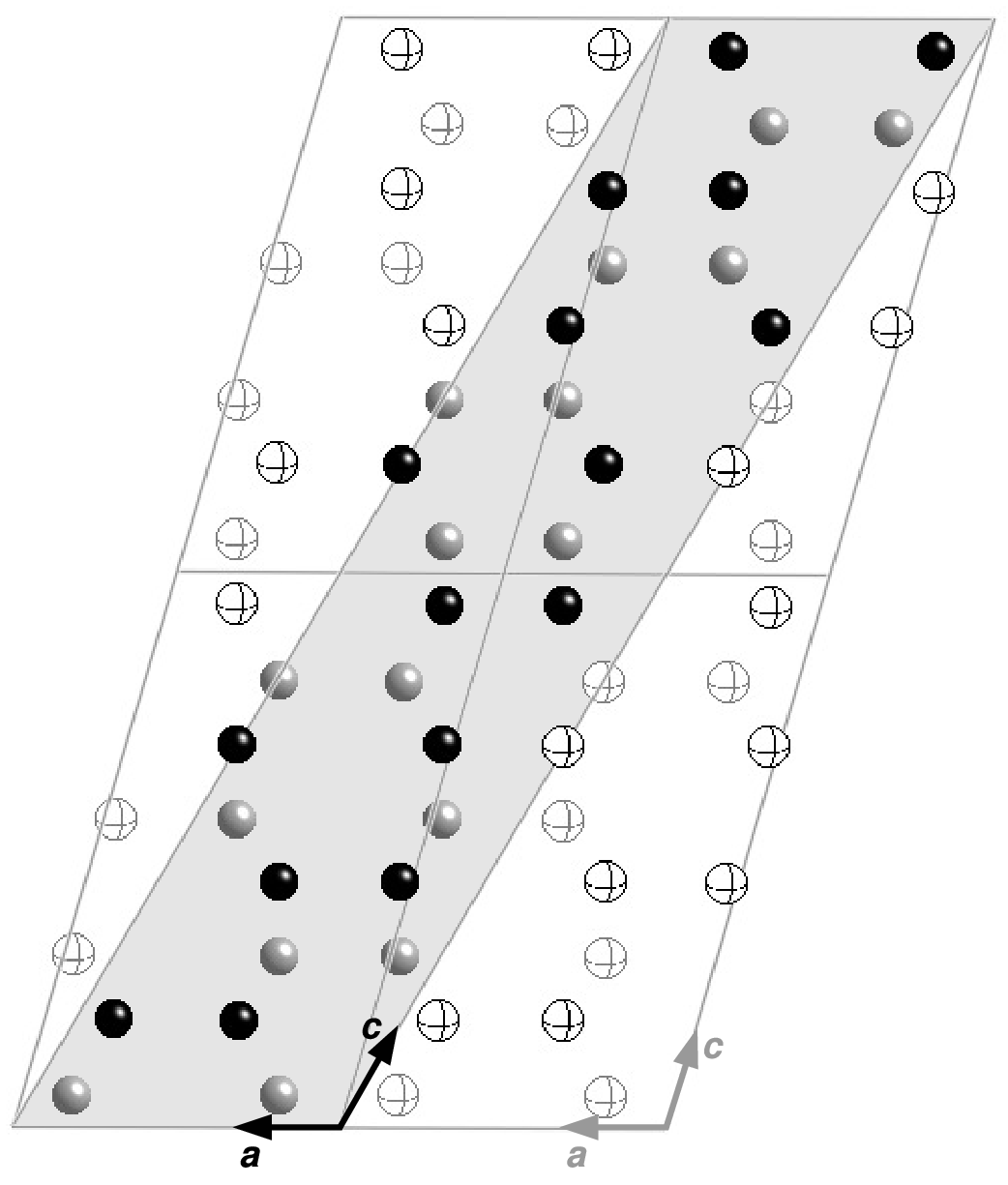}
  \caption{Four conventional monoclinic cells with 16 atoms and an alternative supercell with 32
    atoms (filled) both of which represent the monoclinic crystal. The angles of the two cells are
    $106.3^\circ$ (16-atom cell) and $120.3^\circ$ (32-atom cell). The atoms in the two monoclinic
    planes are drawn as black and gray; the empty circles correspond to the atoms that are present
    in the four 16-atom cells but lie outside of the 32-atom cell.}
  \label{fig_mcl-32atoms}
\end{figure}

When investigating the hcp $\rightarrow$ monoclinic transformation path, we assumed that a 16-atom
sh (hcp) cell transforms to the conventional monoclinic unit cell with 16 atoms and the monoclinic
angle $106.3^\circ$. However, the choice of the tiling of the monoclinic lattice is not unique and
one can construct an infinite number of supercells that represent it, although with larger number of
atoms. For example, it has been shown \cite{hirth:06} that one can construct a supercell with 32
atoms (see \reffig{fig_mcl-32atoms}) whose monoclinic angle is, in our case, $120.3^\circ$. This
angle is very close to that of the initial hexagonal (sh or hcp) lattice and, therefore, basically
no strain is associated with changing the shape of the simulated cell during the transformation. One
would thus expect that using this 32-atom monoclinic supercell, the energy barrier for the sh (hcp)
$\rightarrow$ monoclinic transformation would be lower than that for the conventional 16-atom
monoclinic cell. We performed the simulations above also using this larger 32-atom cell and observed
only minor changes in the variation of energy and volume for the sh $\rightarrow$ monoclinic
transformation. This is shown by comparing the solid and dashed lines in the right panel of
\reffig{fig_esurf} that correspond to the 16-atom and 32-atom cells, respectively. In both cases of
the 32-atom calculation the activation energy decreased relative to the equivalent 16-atom
simulation by an additional 0.05~eV/atom. This brings the total energy barrier for the fcc
$\rightarrow$ sh $\rightarrow$ monoclinic transformation down to about 0.25~eV/atom and that of the
fcc $\rightarrow$ hcp $\rightarrow$ monoclinic transformation to 0.15~eV/atom.

\begin{figure}[htb]
  \centering
  \footnotesize
  \includegraphics[scale=0.25]{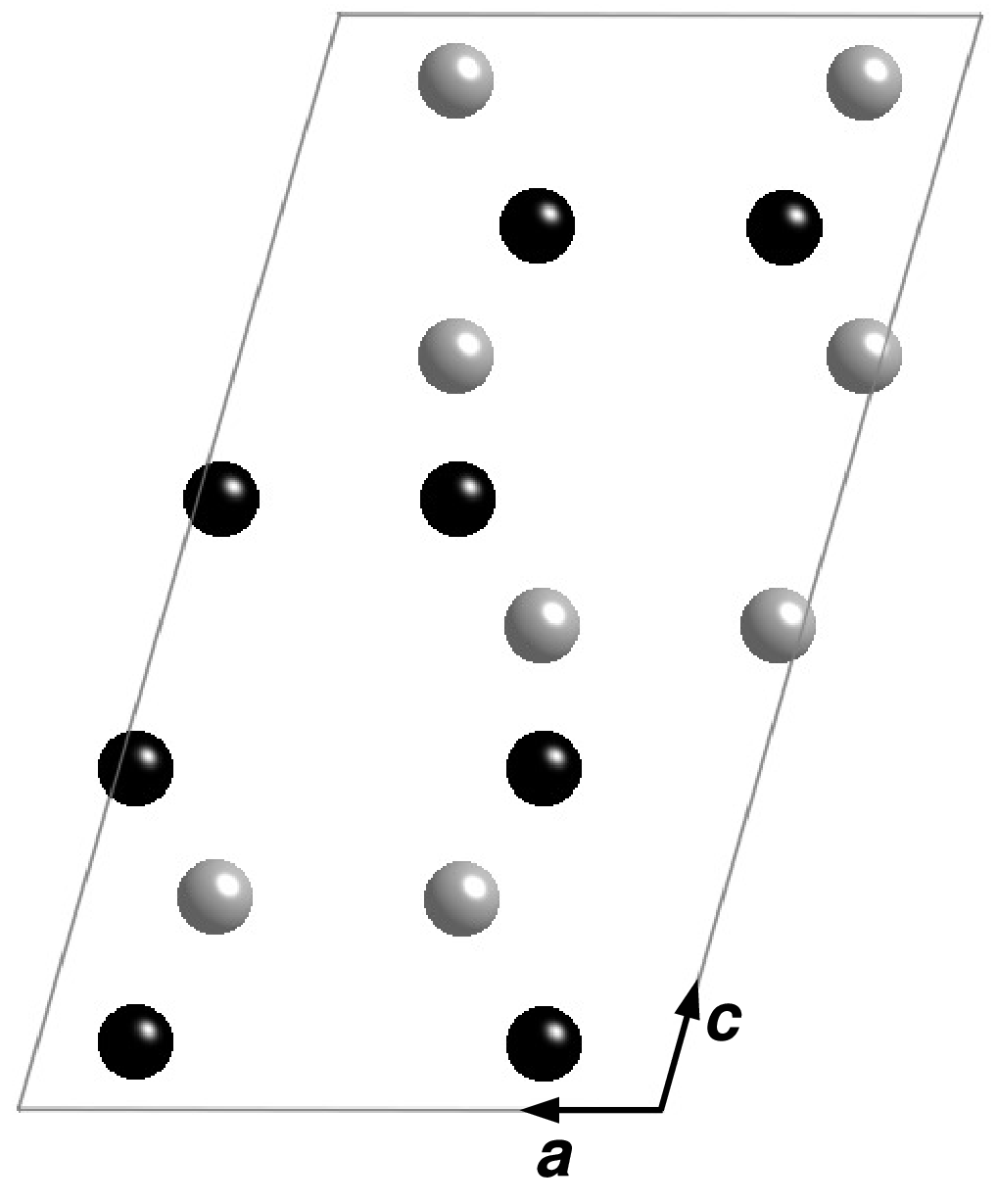} \quad
  \includegraphics[scale=0.25]{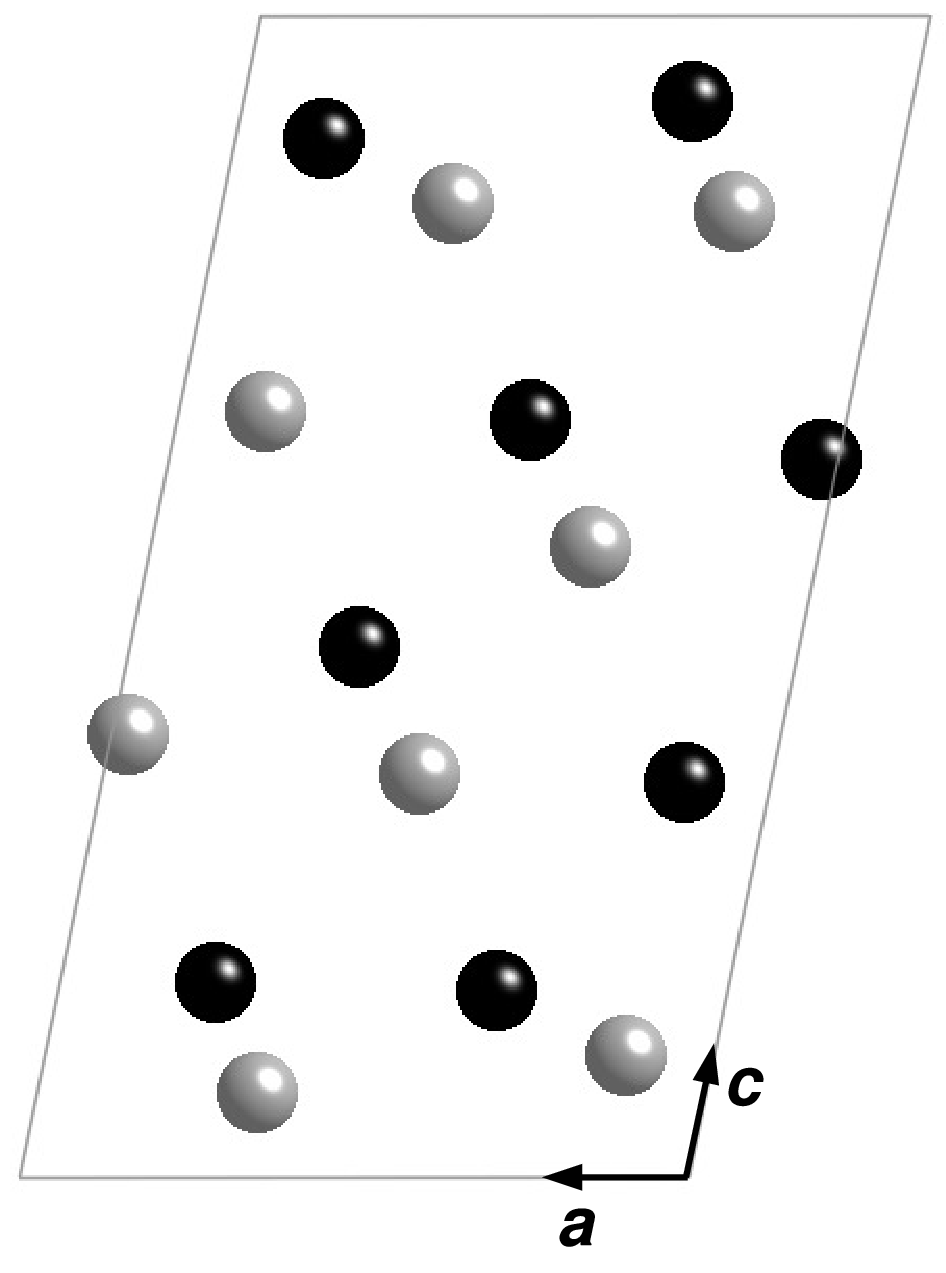} \\
  \footnotesize
  (a) \hskip4cm (b)
  \caption{Atomic positions in the monoclinic $\alpha$ structure calculated using MEAM (a) and those
    deduced from the X-ray powder diffraction measurements of \cite{zachariasen:63a} (b). The same
    colors of atoms belong to the same monoclinic plane in the $\bsym{b}$ direction that is
    perpendicular to the plane of the figure. If the lattice in (a) is rotated around the horizontal
    $\bsym{a}$ axis, the black and gray atoms in the two monoclinic planes overlap and one observes
    a perfect hexagonal pattern that is not present in (b).}
  \label{fig_mcl-compare}
\end{figure}

In order to understand the limitations of the calculations presented in this paper, it is important
to analyze the accuracy with which the MEAM potential represents the individual solid phases of
Pu. The reference phase that is used to construct this potential is the fcc $\delta$ phase for which
the elastic constants \cite{taylor:68, migliori:02, ledbetter:08}, lattice parameters and the
cohesive energy \cite{baskes:00} are well known. The theoretical formalism of any interatomic
potential captures only certain relevant aspects of the physics of bonding. The potential is
typically constructed to reproduce experimental or first principles data for a simple crystal
structure (often cubic) and is tested for transferability to other atomic environments. In the case
of the MEAM potential for Pu, one assumes that parametrizing this potential to the $\delta$ phase
will provide an empirical description that will reproduce correctly the phases whose symmetry and
volume are comparable with those of the reference structure. Indeed, it has been shown in
\cite{baskes:00} that this potential also predicts correct energies of the orthorhombic $\gamma$
and bcc $\epsilon$ phases. It is, however, much more challenging to achieve the same degree of
accuracy for the low-symmetry simple monoclinic $\alpha$ and body-centered monoclinic $\beta$
phases whose volumes are 20\% and 10\% lower, respectively, than the reference fcc $\delta$ phase.
To demonstrate this we show in \reffig{fig_mcl-compare}(a) and (b) a comparison of the equilibrium
monoclinic lattice calculated from MEAM and used throughout this paper with that obtained by
Rietveld refinement of the X-ray powder diffraction measurements \cite{zachariasen:63a},
respectively. Clearly, the positions of atoms calculated from the MEAM potental for Pu do not
correspond to the experiment. The most notable feature is an unphysical hexagonal symmetry that can
be seen clearly by rotating the unit cell about the horizontal $\bsym{a}$ axis. The origin of this
discrepancy is presumably the inability of the potential to account for strong directional bonding
arising from the overlap of the $f$ orbitals as the volume of the crystal is decreased and the
symmetry of the parent fcc phase (48 symmetry operations) is broken into 12 variants of the
monoclinic lattice (4 symmetry operations).

The observation that the current MEAM potential for Pu does not represent with reasonable accuracy
the positions of atoms in the $\alpha$ phase of Pu raises serious concerns when calculating the
transformation pathway between the $\delta$ and $\alpha$ phases. The region where MEAM is least
accurate, and thus the results of our simplified studies presented above are least trustworthy,
corresponds to the shaded area in \reffig{fig_esurf}. For this range of transformation coordinates
$x$, the volume and thus also the lattice parameters (not shown here) change rapidly as the sh (hcp)
phase transforms to the trigonal phase and then to the final monoclinic structure. Nevertheless, the
inaccuracy of the MEAM potential in this low-volume region does not affect our main conclusion that
the hcp structure is predicted as the preferred intermediate symmetry through which the $\delta$
phase transforms to the monoclinic $\alpha$ phase.

\section{Conclusions}

Our objective in this paper was to investigate two prototypical transformation mechanisms for Pu
that were identified in \cite{lookman:08} from purely symmetry considerations. Utilizing the
well-known Shoji-Nishiyama relationship $[111]_\delta$ $\parallel$ $[0001]_{\rm hex}$ $\parallel$
$[010]_\alpha$ that constrains the mutual orientations between the individual phases of Pu, Lookman
et al. \cite{lookman:08} suggested that the transformation of the fcc $\delta$ phase to the
monoclinic $\alpha$ phase proceeds via the intermediate trigonal and either sh or hcp
structures. However, since these considerations do not involve any energetics, they cannot determine
which of these hexagonal structures is more energetically favorable.

In the present work, these calculations were performed using the MEAM for Pu \cite{baskes:00} in
which we assumed that the fractional coordinates of atoms change along the shortest paths between
the initial and the final crystal structures. The individual phonon modes along the transformation
pathway correspond to displacements of atoms in different directions but the vector sum of these
individual vectors yields a straight path between the positions of atoms in their parent and product
phases. The corresponding shape of the unit cell that minimizes the total energy is determined while
keeping the fractional coordinates of atoms fixed. These calculations are performed in the
isothermal-isobaric ensemble at 0~K. We have shown that the current MEAM potential for Pu predicts
that the $\delta$ $\rightarrow$ $\alpha(\alpha')$ transformation occurs preferably via the
intermediate hcp phase. If one uses the 32-atom supercell for the hcp $\rightarrow$ monoclinic
transformation, the total energy barrier is about 0.15~eV/atom and the 20\% volume decrease occurs
entirely during the transformation from the intermediate hcp to the final monoclinic structure. The
existence of an intermediate hcp phase has been observed recently during the pressure-induced
$\delta$ $\rightarrow$ $\alpha'$ phase transformation \cite{faure:06}. For completeness, the
signatures of the intermediate trigonal or hcp symmetries should also be probed with picosecond
resolution by, for example, transient X-ray diffraction techniques that are becoming increasingly
important in many areas of materials science.

We argue that while the MEAM potential for Pu reproduces faithfully the crystal structures whose
symmetries and volumes are comparable with those of the reference fcc $\delta$ phase, its current
formulation cannot capture the positions of atoms in the low-symmetry $\alpha$ and $\beta$ phases of
Pu that were determined earlier by Rietveld refinement of X-ray powder diffraction data
\cite{zachariasen:63, zachariasen:63a}. Hence, also the energetics of the pathways from the
hexagonal intermediate symmetry to the final monoclinic structure is not faithfully reproduced by
MEAM. This makes even this state-of-the-art interatomic potential inappropriate for use in the
methods that search for minimum energy paths between two well-defined configurations, such as the
NEB method \cite{jonsson:98}. 

Continued progress in this field relies on our future ability to understand and at least
qualitatively correctly capture the core physics associated with strong correlations among $f$
electrons. Having an interatomic potential that would successfully reproduce all six crystal phases
of Pu and the seventh under pressure would undoubtably provide new understanding of the physics of
this important metal. This potential will have to reproduce the largest set of requirements ever
imposed on an interatomic potential of a single element. In the case of Pu, these are mainly the
energies and volumes of all crystal structures, softening of the $\langle111\rangle$ transverse
acoustic phonon at the L point in the $\delta$ phase that drives the phase transformation to the
$\alpha'$ phase, and positions of atoms in the low-symmetry $\alpha$ and $\beta$ structures. One
promising candidate is the modified generalized pseudopotential method (MGPT) developed by Moriarty
et al. \cite{moriarty:06} and successfully applied to Mo and Ta. Recent advances of this method
allowed simulations of $f$ electron materials such as $\delta$ Pu in which the strong correlations
are modeled by turning off the bonding due to the $f$ electrons completely and describing the
material as a $d$ band metal. Unlike the MEAM potential for Pu, this MGPT potential has been
demonstrated to capture reasonably well the phonon dispersion curves, although the anomalous
softening of the transverse $\gdir{111}$ phonon at the L point is not reproduced. Another possible
candidate for the construction of a better interatomic potential is the tight-binding-based
description within the bond order potential (BOP) \cite{horsfield:96} that has been shown to be
imminently suitable for transition metals crystallizing in the bcc structure with approximately
half-filled $d$ bands (V, Nb, Mo, Ta, W) and $\alpha$ Fe. Although this BOP is not currently
available, the first step was undertaken by Hachiya \cite{hachiya:99} who constructed a
tight-binding description of fcc $\alpha$ phase of Th and $\delta$ Pu, both of which are
characterized by approximately half-filled $f$ orbitals. However, more extensive tests of this
approach, in particular its ability to describe the space group of the monoclinic $\alpha$ phase of
Pu and the large volume changes between the $\delta$ and $\alpha$, $\beta$ phases of Pu is still
lacking.

\section*{Acknowledgments}

We are grateful to Jim Smith for his inspiration and interest in our work on actinides and phase
transformations.  The authors acknowledge many discussions on the topic with Mike Baskes, Jason
Lashley, Steve Valone, Chris Taylor, Terry Mitchell and Siegfried Hecker. This work has been
supported by the LDRD project of the Seaborg Institute for Transactinium Science, Los Alamos
National Laboratory.

\newpage
\bibliography{bibliography}

\end{document}